\documentclass[prb,aps,superscriptaddress,twocolumn,floatfix,10pt]{revtex4-1}
\usepackage{graphicx,color}
\usepackage{amsmath,amssymb,bm}
\usepackage{braket,dsfont}
\usepackage{bbold}
\usepackage{hyperref}

\newcommand{\HS}{\mathcal{H}}
\DeclareMathOperator{\tr}{Tr}
\newcommand{\id}{\mathbb{1}}
\newcommand{\im}{\mathrm{i}}

\newcommand{\Renyi}{R\'enyi }
\newcommand{\Uanc}{U_{\textrm{anc}}}
\newcommand{\Udis}{U_{\textrm{disent}}}
\newcommand{\betamax}{\beta_{\textrm{max}}}
\newcommand{\entropy}{S_{LL'}}
\newcommand{\entropyopt}{\widetilde{E}_p}

\begin{document}
\title{Finding purifications with minimal entanglement}

\author{Johannes Hauschild}
\email[E-mail: ]{johannes.hauschild@tum.de}
\affiliation{Department of Physics, T42, Technische Universit{\"a}t M{\"u}nchen, James-Franck-Stra{\ss}e 1, D-85748 Garching, Germany}
\author{Eyal Leviatan}
\affiliation{Department of Condensed Matter Physics, Weizmann Institute of Science, Rehovot 7610001, Israel}
\author{Jens H. Bardarson}
\affiliation{Department of Physics, KTH Royal Institute of Technology, Stockholm SE-10691, Sweden}
\author{Ehud Altman}
\affiliation{Department of Physics, University of California, Berkeley, CA 94720}
\author{Michael P. Zaletel}
\affiliation{Department of Physics, Princeton University, Princeton, NJ 08540, USA}
\author{Frank Pollmann}
\affiliation{Department of Physics, T42, Technische Universit{\"a}t M{\"u}nchen, James-Franck-Stra{\ss}e 1, D-85748 Garching, Germany}

\begin{abstract}
Purification is a tool that allows to represent mixed quantum states as pure states on  enlarged Hilbert spaces.
A purification of a given state is not unique and its entanglement strongly depends on the particular choice made.
Moreover, in one-dimensional systems, the amount of entanglement is linked to how efficiently the purified state can be represented using matrix-product states (MPS).
We introduce an MPS based method that allows to find the minimally entangled representation by iteratively minimizing the second \Renyi entropy.
First, we consider the thermofield double purification and show that its entanglement can be strongly reduced especially at low temperatures.
Second, we show that a slowdown of the entanglement growth following a quench of an infinite temperature state is possible.
\end{abstract}

\maketitle

\section{Introduction}
Simulating quantum many-body systems faces a fundamental difficulty due to the complexity required to represent highly entangled states.
Significant progress has been made through the observation that quantum ground states of interest often have only limited (area-law) entanglement, and thus can be represented efficiently using matrix-product states (MPS) \cite{Fannes-1992,Hastings2006,Schollwock2011} in one dimension (1D) and generalized tensor-product states \cite{Verstraete2004PEPS} in higher dimensions.
Such approaches have been particularly successful in the study of ground-state properties of 1D systems, where the density matrix renormalization group (DMRG)  method \cite{White1992DMRG} revolutionized the efficiency of numerical methods.

To extend the success of DMRG to transport and non-equilibrium phenomena, it is necessary to simulate real-time evolution \cite{Vidal2004TEBD,Daley2004tDMRG,White2004tDMRG}.
The bipartite entanglement of pure states generically grows linearly with time, which leads to a rapid exponential blow up in computational cost, limiting pure-state time evolution to rather short times.
But, while the entanglement growth limits the ability to compute the real time evolution of pure quantum states, it need
not impose the same restriction on the imaginary time evolution of mixed states \cite{Barthel2017,Dubail2017}.
It is then natural to ask if the time evolution of mixed states can be represented efficiently using MPS and what sets the difficulty of such computations.

\begin{figure}
	\includegraphics[width=\columnwidth]{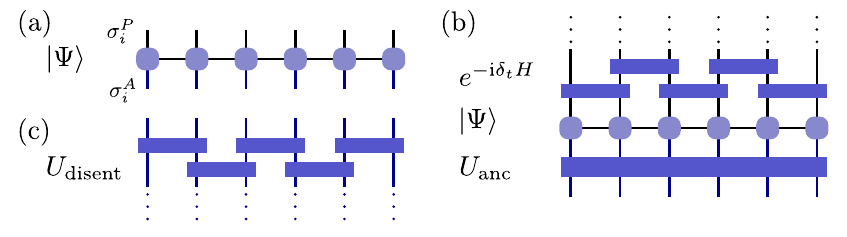}
	\caption{(a) Schematic representation of purified states using MPS.
		(b) A purified state is evolved in real or imaginary time by acting on the physical degrees of freedom (e.g., using a Trotter decomposition of the time-evolution operator).
		The auxiliary degrees of freedom
		are only defined up to a global unitary $\Uanc$ which can be chosen to minimize the entanglement on the bonds.
		(c) The global $\Uanc$ is decomposed into a network of two-site gates to produce a disentangler $\Udis$.}
    \label{fig:disentanglers}
\end{figure}

There are different techniques for simulating mixed states using MPS methods, including
a direct representation of the density matrix as a matrix product operator (MPO) \cite{Zwolak2004},
using minimally entangled typical thermal states (METTS) \cite{White2009,Stoudenmire2010,Binder2017,Berta2017},
and purification \cite{Verstraete2004Purification,Barthel2009}; in this paper we focus on the latter.
In purification, a density matrix $\rho$ acting on a {\em physical} Hilbert space $\HS^P$ is represented as a pure state $\ket{\psi}$ in an enlarged space $\HS^P \otimes \HS^A$:
\begin{equation}
	\rho  = \tr_A \ket{\psi}\bra{\psi}.
	\label{eq:pure}
\end{equation} 
It is always sufficient to choose $\HS^A$ to be identical to $\HS^P$, ``doubling'' each degree of freedom (DoF) as illustrated in  Fig.~\ref{fig:disentanglers}(a).
We note that the purification description can be a limitation for infinite systems \cite{deLasCuevas2013,deLasCuevas2016}. 
Yet on finite systems, a purification can be found formally by diagonalizing the density matrix. In equilibrium this gives the thermofield double (TFD) purification,
$\ket{\psi_\beta} = \frac{1}{\sqrt{Z}} \sum_n e^{- \beta E_n / 2} \ket{n}_P \ket{n}_A$, where $\ket{n}$ are the eigenvectors and $E_n$  the eigenvalues of the Hamiltonian.
It was recently argued that the TFD state can be efficiently represented with an MPS of bond dimension that grows at most polynomially with the inverse temperature \cite{Barthel2017}.
The TFD is only one possible choice of purification, since Eq.~\eqref{eq:pure} is left invariant under an arbitrary
unitary transformation $\Uanc$ which acts only on the ancilla space $\HS^A$.

This gauge freedom may be used to reduce the entanglement in $\ket{\psi}$, rendering the MPS representation more efficient \cite{Karrasch2012,Barthel2013}.
Here, we propose a way to find the minimally entangled purification.
This minimum defines the entanglement of purification $E_p$ \cite{Terhal2002} [defined below in Eq.~(\ref{eq:EP})],
which thus plays a role similar to the entanglement entropy in the pure case: it bounds the bond dimension $\chi \geq e^{E_p}$ \cite{Schuch2008}.
However, this lower bound is irrelevant unless there is an efficient algorithm to \emph{find} the minimally entangled purification at a cost comparable to DMRG [e.g., $\mathcal{O}(\chi^3)$],
which, since it constitutes a global optimization problem over the many-body Hilbert space, is not \emph{a priori} obvious.

Below we introduce a method to find an approximately optimal purification by sequentially applying local disentangling operations to the ancilla DoF.
The cost of the disentangling procedure is comparable to DMRG, and the resulting entanglement $\entropyopt$ reproduces the known properties of $E_p$ in certain limits.
We use the method to optimize both the equilibrium purification and that of a time-dependent state.
We find that the method can significantly slow the entropy growth during real-time evolution down,
as we demonstrate for both the transverse field Ising model and a disordered Heisenberg chain.
For the latter, we find a slow spreading of $\entropyopt$ already for intermediate disorder strengths.
In equilibrium, $\entropyopt$ approaches half of the entropy in the TFD state at low temperature.

Before proceeding we comment on the difference between the method presented here and two other proposals to compute long-time dynamics efficiently using MPS.
Some of us have shown recently that the dynamics of local quantities in thermalizing systems can be captured accurately using the time-dependent variational principle (TDVP) \cite{Leviatan2017}, allowing to extract transport coefficients and even characteristics of chaos.
Another one of us proposed a new truncation method to approximate the time evolution of a density matrix, represented as an MPO, to long times \cite{White2017}.
Both of these methods rely on the assumption that the increase of the non-local information encoded by the ever-growing entanglement entropy is irrelevant to the evolution of observable properties in thermalizing systems.
These methods attempt to simulate the correct macro-state rather than the nearly exact microstate.
Thus, the ``truncation error'' as usually defined in DMRG studies can be large as it is measured with respect to the exact state.
In contrast, the approach presented here attempts, by optimizing the purification, to minimize the truncation error in order to compute the exact micro-state.

\section{Purifications within the MPS formalism}

To represent a purification as an MPS, we take $\HS^P \sim \HS^A$ so that each ``site'' contains a doubled DoF.
The purification then takes the standard MPS form with an enlarged local Hilbert space as shown in Fig.~\ref{fig:disentanglers}(a).
At infinite temperature, the TFD purification is obtained by maximally entangling the physical and ancilla DoF on every site $i$, e.g.,
$\ket{\psi_{0}} = \prod_i \left( \frac{1}{\sqrt{d}} \sum_{\sigma_i} \ket{\sigma_i}_P \ket{\sigma_i}_A \right)$, where $\sigma_i$ runs over the local Hilbert space, resulting in a $\chi = 1$ MPS.
In the standard purification approach, the finite-temperature TFD is obtained from $\ket{\psi_0}$ using imaginary-time evolution,  $\ket{\psi_\beta} \propto e^{-\frac{\beta}{2}H} \ket{\psi_0}$, 
from which thermal expectation values are evaluated as $\braket{Y }_\beta  = \braket{\psi_\beta| Y |\psi_\beta}$. Here, $H$ acts only on $\HS^P$.
Similarly, to compute dynamical properties, for instance, $C(t, \beta) = \braket{B^\dagger Y(t) B}_\beta$, we define $\ket{B(t,\beta)} = e^{- \im t H} B \ket{\psi_\beta}$, so that $C(t, \beta) = \bra{B(t,\beta)} Y \ket{B(t,\beta)}$.
By taking $B = e^{\im \epsilon X}$, this form is sufficient to find quantities of interest such as the spectral function $-\im \partial_\epsilon C(t, \beta) = A_{YX}(t, \beta) = \braket{[Y(t), X(0)]}_\beta$.
The requisite time evolution (both imaginary and real) can be  simulated using standard methods \cite{Vidal2004TEBD,Daley2004tDMRG,White2004tDMRG,Haegeman2011,Zaletel2015}.

The computational complexity of such simulations is generically linked to the bipartite von-Neumann entanglement entropy $\entropy = - \tr(\rho_{LL'} \log(\rho_{LL'}))$,
where $\rho_{LL'} = \tr_{RR'}(\ket{\psi}\bra{\psi})$ is the reduced density matrix defined by a bipartition $\HS^P = L \otimes R $ and $\HS^A = L' \otimes R'$ at any of the bonds of the MPS; 
the bond dimension $\chi$ is bounded by $ \chi \geq e^{\entropy}$.
Since other purifications can be obtained by acting with $\Uanc$ on the ancilla space, see Fig.~\ref{fig:disentanglers}(b), 
it is desirable to exploit this choice to reduce $\entropy$.
Karrasch \emph{et al.}~\cite{Karrasch2012} noticed that a natural choice is the ``backward time evolution,'' $\Uanc = e^{\im t H}$,
because if $B$ is local, this choice leaves $\ket{B(t,\beta)}$ invariant outside the growing ``light cone'' of the perturbation.
Barthel~\cite{Barthel2013} improved this approach by evolving both $X$ and $Y$ in the spectral function
$A_{YX}(t, \beta) = \braket{[Y(t), X(0)]}_\beta = \braket{[Y(t/2), X(-t/2)]}_\beta$ as Heisenberg operators, 
which allows reaching times twice as long with comparable numerical effort \cite{Barthel2013,Karrasch2013}.
However, these prescriptions need not be optimal; ideally, we would minimize $\entropy$ over all possible purifications, which would result in the entanglement of purification $E_p$ \cite{Terhal2002}:
\begin{equation}
E_p[\rho_{LR}] \equiv \min_{\ket{\psi}} \entropy\left[\ket{\psi}\right] = \min_{\Uanc} \entropy\left[\Uanc\ket{\psi}\right]. \label{eq:EP}
\end{equation}
Equivalently, given an ansatz purification $\ket{\psi}$, we search for $\Uanc$ such that $\Uanc \ket{\psi}$ has minimal entanglement;
from this perspective, $\Uanc$ is a ``disentangling'' operation.

\section{Disentangling algorithm}
We propose an algorithm to approximately identify the optimal $\Uanc$ via a sequence of local disentangling operations, producing a circuit $\Uanc = \Udis$ of the form shown in Fig.~\ref{fig:disentanglers}(c):
The time evolution is applied to the purified state using the time-evolving block-decimation algorithm (TEBD) \cite{Vidal2004TEBD}.
The TEBD algorithm \cite{Vidal2004TEBD} is based on a Trotter decomposition of $e^{-\im t H}$ into two-site unitaries
$e^{-\im \delta_t H_{i,i+1}}$ as illustrated in Fig.~\ref{fig:disentanglers}(b).
These unitaries are applied to the physical indices of the effective two-site wave function
\begin{equation*}
\ket{\Theta} = \sum_{\substack{\sigma^P_i, \sigma^A_i, l \\ \sigma^P_{i+1}, \sigma^A_{i+1}, r}}
\Theta^{\sigma^P_i \sigma^A_i, \sigma^P_{i+1} \sigma^A_{i+1}}_{l,r} \ket{l} \ket{\sigma^P_i\sigma^A_i} \ket{\sigma^P_{i+1}\sigma^A_{i+1}} \ket{r},
\end{equation*}
where $\ket{l}$ (and $\ket{r}$) labels a basis consisting of Schmidt states to the left of site $i$ (and right of site $i+1$, respectively).
During a real-time evolution, we disentangle the two-site wave function right after each Trotter step using a unitary acting on the auxiliary space.
These two-site disentanglers can be found using an iterative scheme based on minimizing the second \Renyi entropy as a cost function as explained below,
similar to the optimizations of a multi-scale entanglement renomormalization ansatz (MERA) \cite{Evenbly2014}.
As the time evolution proceeds, the disentangling unitary circuit $\Udis$ is then gradually built up by two-site unitaries, as depicted in Fig.~\ref{fig:disentanglers}(c).
During an imaginary-time evolution we use a different scheme outlined in Sec.~\ref{sec:global-disent}.

While the algorithm can suffer from numerical instabilities, we find empirically that it converges to a purification
with significantly less entanglement compared to both backward time evolution and no disentangling at all, as shown in the benchmark section.
The method described above is particularly suitable for correlation functions which involve only a single purification, e.g., $C(t, \beta)$, as there is no need to keep track of $\Udis$. 
When two distinct purifications $\ket{B(t)}$ and $\ket{A(t)}$ are required, one would have to compress ${\Udis}^\dagger_B {\Udis}_A$ as a separate MPO.

\begin{figure}
	\includegraphics[width=\columnwidth]{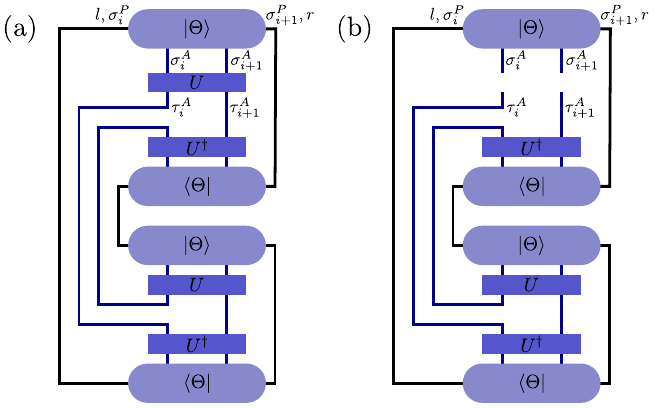}
	\caption{
		(a) Tensor network for $Z_2(U, \Theta)$. 
		(b) Effective environment $E_2(U,\Theta)$ such that $Z_2(U, \Theta) = \tr\left(U\,E_2(U,\Theta)\right)$.
	}
\label{fig:network}
\end{figure}

\subsection{Two-site disentangler minimizing the entropy}
\label{sec:entropydisentangler}

We explain now how to find a two-site unitary $U = U^{\tau^A_i,\tau^A_{i+1}}_{\sigma^A_i,\sigma^A_{i+1}}$ (i.e.,
acting in $\HS^A$) which minimizes the entanglement of an effective two-site wave function $U \ket{\Theta}$, similar as
during the optimzation of MERA \cite{Evenbly2014}.
We chose to minimize the second \Renyi entropy $ S_2(U \ket{\Theta}) = - \log \tr\left(\rho_{LL'}^2\right)$,
where $\rho_{LL'} $ is the reduced density matrix  $\rho_{LL'} = \tr_{\sigma^P_{i+1},\tau^A_{i+1},r} \left(U\ket{\Theta}\bra{\Theta}U\right)$ \cite{Nielsen2000}.
In contrast to the von-Neumann entropy, the second \Renyi entropy is readily expressed as $S_2(U\ket{\Theta}) = -\log(Z_2)$ with the tensor network $Z_2$ depicted in Fig.~\ref{fig:network}(a); $Z_2$ is to be maximized.
We solve this non-linear optimization problem iteratively: 
in the $n$-th iteration, we consider one $U_{n+1}$ formally as independent of the other $U_n$ and write  $Z_2(U_{n+1},U_n, \Theta) = \tr\left(U_{n+1}\,E_2(U_n,\Theta)\right)$,
where the network for the ``environment'' $E_2(U_n,\Theta)$ is shown in Fig.~\ref{fig:network}(b).
It is easy to see that the unitary $U_{n+1}$ maximizing this expression is given by a polar decomposition of $E_2(U_n,\Theta)$, in other words we set $U_{n+1} := Y X^\dagger$ where $X$ and $Y$ are obtained from a singular value decomposition of $E_2(U_n, \Theta) = X \Lambda Y^\dagger$.
The unitary minimizing $Z_2(U,\Theta)$ is then a fixed point $U_{*}$ of this iteration procedure.
As a starting point of the iteration, one can choose the identity $U_1 := \id$.
At later times, one can also use the result of $U_n$ from previous iterations (for the same time step and at the same bond) as initial guess for the next disentangler, which reduces the number of necessary iterations in many cases.

Since this iteration is based on a descent, it tends to go into local minima within the optimization space.
To find the global optimum, we can perform multiple iterations in parallel:
one starting from the identity, and others starting from initially random unitaries (chosen according to the Haar
measure, i.e., from the so-called circular unitary ensemble).
From the unitaries obtained by the parallel iterations, we choose the one with the smallest final entropy.

The disentangler $U_n$ obtained by the above procedure preserves the quantum numbers of symmetries in the Hamiltonian,
at least if the initial guess $U_0$ preserves them.
In the presence of such a symmetry one should choose $U_0$ accordingly from the Haar measure on unitaries preserving the
symmetry to avoid an artificial build-up of entanglement.
In our case, we exploited the $S^z$ conservation in the Heisenberg chain~\eqref{eq:Heisenberg} to reduce the computational cost in the tensor contractions and singular value decompositions \cite{Singh2010,tenpy}.

\subsection{Two-site norm disentangler}
\label{sec:normdisentangler}
In this section, we discuss an alternative way to obtain a two-site disentangler,
which directly focuses on the required bond dimension. 
The procedure described below is equivalent to finding the ``entanglement branching operator'' introduced by \citet{Harada2018}.
In order to reduce the bond dimension, we look for a two-site unitary $U$ (acting on the ancilla DoF) 
for which the trunction of the effective two-site wave function $U\ket{\theta}$ has the smallest truncation error.
To find this $U$, we use a similar, iterative scheme as above: given $U_n$,
we calculate the truncated $\left(U_n\ket{\theta}\right)_{\mathrm{trunc}}$ and find the $U_{n+1}$ maximizing the overlap 
$\left|\bra{\theta}U_{n+1}^\dagger (U_n\ket{\theta})_{\mathrm{trunc}} \right|$. Again, the new $U_{n+1}$ can
be found by a polar decomposition of the ``environment'' consisting of the corresponding tensor network for
$\left|\bra{\theta}U_{n+1}^\dagger (U_n\ket{\theta})_{\mathrm{trunc}} \right|$, but excluding the $U_{n+1}$.
Since the optimal $U$ depends on the final bond dimension $\chi$ after truncation, we need to gradually increase $\chi$ and
repeat the iteration procedure until the truncation error for the given bond dimension is below a desired accuracy threshold.
While we found that this gradual increase of $\chi$ also helps to find the optimal disentangler,
it substantially increases the computational cost.

\subsection{Global disentangling for imaginary-time evolution}
\label{sec:global-disent}
In contrast to the real-time evolution, the Trotter gate $e^{- \delta_\beta H_{i,i+1}}$ in imaginary-time evolution is non-unitary.
Thus, it can change the Schmidt values and thus generate entanglement on sites it does not even act on, which creates
the necessity for a more global scheme of disentangling than the one presented above for the real-time evolution.
Instead, we perform the imaginary time evolution as usual (with $\Uanc = \id$) and disentangle only after each $n$th time step in a more global fashion:
in this case, we find that generating the network of $\Udis$ by optimizing bonds with right and left sweeps similar as in DMRG
is more effective than the Trotter-type scheme of even and odd bonds depicted in Fig.~\ref{fig:disentanglers}(c).
Moreover, it is straight-forward to generalize the two-site disentangling described above to multiple sites by grouping
multiple sites.
For example, we can disentangle the wave function of four sites $i, i+1, i+2, i+3$ by grouping each two sites as $(i, i+1)$ and $(i+2, i+3)$  and then using the above-described method.
As the resulting disentangler can perform arbitrary ``on-site'' rotations within each group, it is necessary to disentangle the obtained wave function (recursively) within each group.
While such a grouping provides additional freedom in the unitary to be found and is thus a systematic improvement for
finding the optimal \emph{global} disentangler,
it comes at the cost of a scaling of required computational resources which is exponential in the number of included sites.
In practice, we limited ourselves to optimizing at most four sites at once.

As an alternative for the global disentangling, we tried a method along the lines of \citet{Hyatt2017}.
Here, the idea is to identify pairs of sites with maximal mutual information as candidates for disentangling.
Using swap gates (commonly used for TEBD with longer-range interactions \cite{Stoudenmire2010}), we bring the two sites next to each other and disentangle them with a two-site
disentangler as described above for the real-time evolution.
Yet, we find that this approach is very limited by the fact that the purification can not be disentangled completely
(except for $\beta\rightarrow \infty$), such that we fail at some point to identify the next candidate pair to be disentangled.

\section{Benchmarks}
\label{sec:benchmark}
\subsection{Finite temperatures}
To benchmark our algorithm, we study a concrete example,
the generalized transverse field Ising model
\begin{equation}
H = - J^x \sum_{i=1}^{L-1} \sigma^x_i \sigma^x_{i+1} -  J^z \sum_{i=1}^{L-1}  \sigma^z_i \sigma^z_{i+1}
    - h^z \sum_{i=1}^{L} \sigma^z_i.
\label{eq:Ising}
\end{equation}
For $J^z = 0$, the model maps onto free fermions and exhibits a quantum phase transition at $h^z_c = J^x$.
The term proportional to $J^z$ introduces interactions and breaks integrability.

\begin{figure}
    \includegraphics[width=\columnwidth]{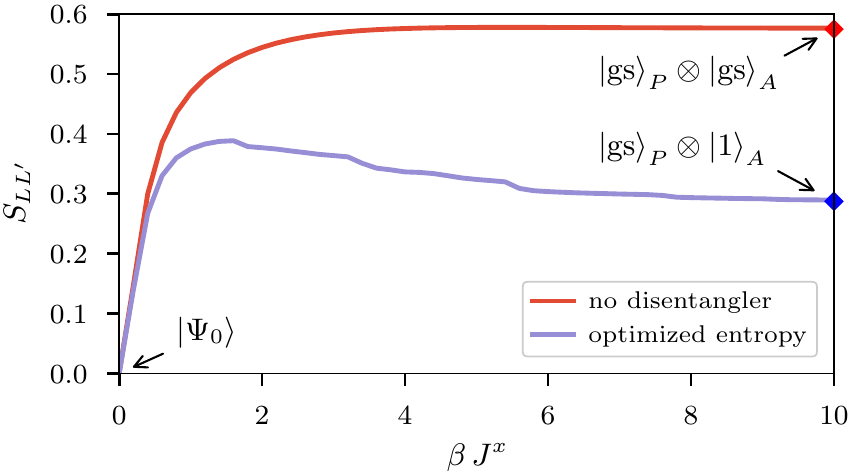}
    \caption{Half-chain entanglement entropy of the finite-temperature purification $\ket{\psi_\beta} \propto
		e^{-\frac{\beta}{2}H} \ket{\psi_0}$ in the generalized Ising model \eqref{eq:Ising} with $L=50$ sites, for the
		TFD state ($\Uanc=\id$, upper line) and when disentangling up to four sites at once (lower line).
    The parameters $J^x = h^z=1$ and $J^z=0.1$ are chosen to be in the vicinity of the quantum phase transition.
    The diamonds on the right axis indicate the half-chain entanglement $S_\mathrm{gs}$ (blue) and $2 S_\mathrm{gs}$ (red) of the ground state $\ket{\mathrm{gs}}$ obtained from DMRG.
    }
    \label{fig:Ising_finiteT}
\end{figure}

Figure~\ref{fig:Ising_finiteT} compares the entanglement of the optimized purification with the entanglement of the TFD state obtained by imaginary time evolution without disentangling, i.e., $\Uanc=\id$.
The infinite temperature state $\ket{\psi_0}$ has maximal entanglement between the physical and auxiliary DoF on each site, but no correlations between different sites, hence $\entropy=0$.
For small $\beta$, the imaginary time evolution starts to build up correlations between neighboring sites, but it is not immediately possible to disentangle the state with a rotation in $\HS^A$.
For example, a non-trivial unitary acting on $\sigma^A_i$ and $\sigma^A_{i+1}$ would lead to a strong correlation between $\sigma_i^P$ and $\sigma^A_{i+1}$, and thus larger entanglement for a cut between sites $i$ and $i+1$.
However, due to the monogamy of entanglement, the build-up of quantum correlations between different sites ensures the reduction of the entanglement between the physical and auxiliary spaces.
Consequently, the disentangler can reduce the entanglement at larger $\beta$.
This is most evident in the limit of large $\beta$ in which $e^{-\frac{\beta}{2} H} $ becomes a projector $\ket{\mathrm{gs}}\bra{\mathrm{gs}}$ onto the ground state $\ket{\mathrm{gs}}$.
In this limit, the TFD purification ends up with two copies $\ket{\mathrm{gs}}_P \otimes \ket{\mathrm{gs}}_A$ of the ground state in $\HS^P$ and $\HS^A$.
In contrast, a perfect disentangling algorithm should be able to rotate $\ket{\mathrm{gs}}_A$ into an unentangled
product state $\ket{1}_A$, ending up with the state $\ket{\mathrm{gs}}_P \otimes \ket{1}_A$ which has only half as much entanglement as the TFD.
The fact that we find a purification with an entanglement close to that of the ground state shows that our algorithm can indeed find the minimum, i.e., it finds $E_p$.

\begin{figure}
    \includegraphics[width=\columnwidth]{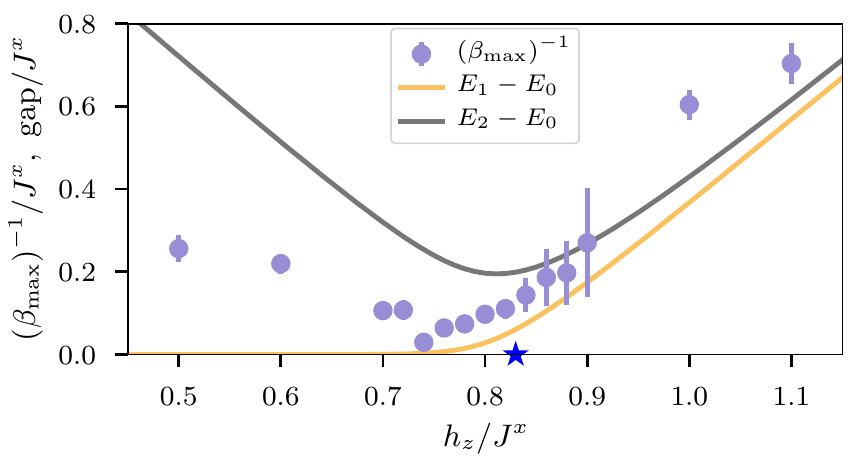}
	\caption{Behavior of the position of the maximum $\betamax$ in Fig.~\ref{fig:Ising_finiteT} with the parameter
		$h^z$, for $L=50$ sites and $J^z = 0.1 J^x$. 
		Error bars indicate uncertainties in extracting $\betamax$ stemming from a limited resolution in $\beta$ and numerical instabilities of the algorithm.
		For comparison, the energy gaps of the first and second states above the ground state (extracted with DMRG) are
		also shown.
		The critical $h^z/J^x$ in the thermodynamic limit is indicated by the blue star on the $x$ axis.
    }
    \label{fig:Ising_beta_max}
\end{figure}
Notably, we also find a maximum at intermediate $\beta$ (although our algorithm suffers from numerical instabilities in this region).
This can be understood from the fact that the entanglement of purification has contributions from both quantum fluctuations and thermal fluctuations, and the latter vanish for $\beta \rightarrow \infty$.
A similar maximum is also present in the holographic prescription for the entanglement of purification \cite{Takayanagi2017,Nguyen2017}.
Figure~\ref{fig:Ising_beta_max} shows that the maximum moves to larger $\beta$ when tuning $h^z$ towards the phase transition.
We attribute this increase of $\betamax$ to the closing energy gap which induces thermal fluctuations at smaller
temperatures (and thus additional entanglement entropy in the purified state on top of the
ground-state entropy reached in the limit $\beta \rightarrow \infty$).
In the symmetry-broken phase for $h^z \lesssim 0.75$, the ground state is (for the finite system almost) two-fold
degenerate, and $\entropy(\beta \rightarrow \infty)$ is increased by $\log(2)$ on top of the ground-state entanglement entropy.
We still observe a maximum of $\entropy$ at finite $\beta$ in this phase, yet less pronounced than in the paramagnetic phase.

\begin{figure}
	\includegraphics[width=\columnwidth]{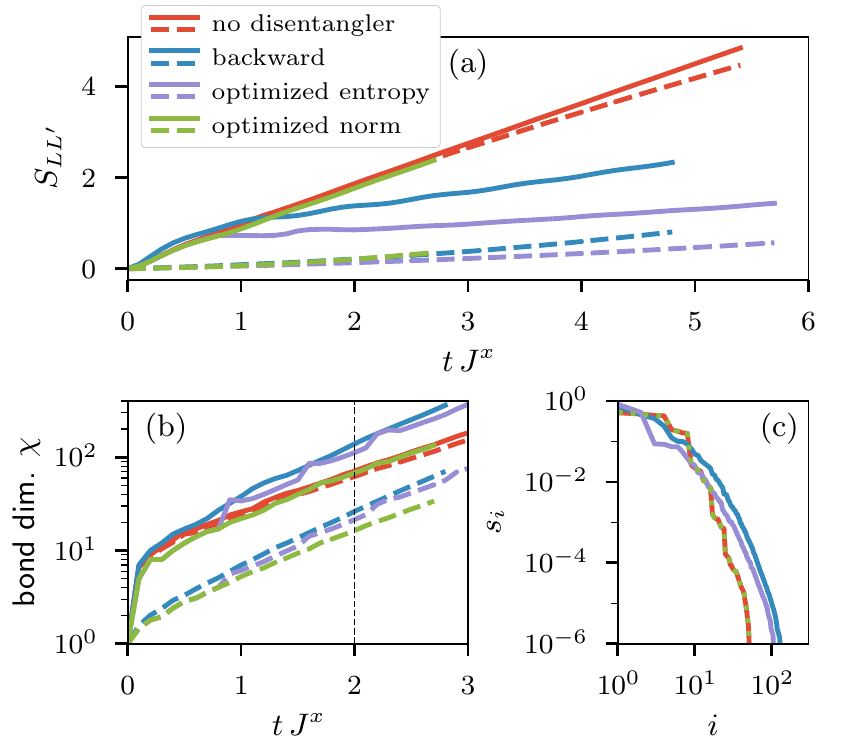}
	\caption{(a) Comparison of the entanglement in the purification state $\ket{S^{+}_{L/2}(t,\beta=0)} = e^{-\im t H} S^{+}_{L/2}\ket{\psi_0}$
		for the Ising chain \eqref{eq:Ising} with $L=40$ sites, $J^x=h^z=1, J^z=0.1$.
		(b) MPS bond dimension when the truncation error is kept below $ 10^{-6}$ in each step.
		(c) Decay of the Schmidt values $s_i$ on the central bond at time $t J^x=2$.
		In panels (a) and (b), solid lines (dashed lines) show the maximum (mean) over different bonds.
		In all panels, different colors compare different disentanglers $\Uanc$.
	}
	\label{fig:Ising_Svst}
\end{figure}

\subsection{Real time evolution at infinite temperature}
Next, we consider the time evolution of a local operator applied to the infinite-temperature purification $\ket{S^+_{L/2}(t,\beta=0)} = e^{-\im t H} S^+_{L/2}\ket{\psi_0}$, 
where $S^+_i = S^x_i + \im S^y_i$.
Figure~\ref{fig:Ising_Svst}(a) compares the resulting entanglement for no disentangling ($\Uanc = \id$), backward time
evolution ($\Uanc = e^{i t H}$), and the optimized disentangler ($\Uanc = \Udis$) using the two-site disentanglers
described in Secs.~\ref{sec:entropydisentangler} and \ref{sec:normdisentangler}. 
Note that for $\beta = 0$ backward time evolution is equivalent to the Heisenberg evolution of $S^+_{L/2}$.
The maximum of the entropy over different bonds (solid lines) grows roughly linear in all three cases, yet with very different prefactors.
While the growth is spatially almost uniform in the case $\Uanc=\id$,
both the backward time evolution and our optimized algorithm develop entropy only within a causal ``light-cone,'' 
which leads to a significant reduction when the mean over different bonds is taken (dashed lines).
Figure~\ref{fig:Ising_Svst}(b) compares the growth of the required MPS bond dimension when the truncation error is kept fixed.
Both backward time evolution and the optimized disentangler minimizing the entropy require a slightly higher maximal bond dimension close to where $S^{+}_{L/2}$ was applied.
This apparent contradiction of a larger bond dimension despite a lower entropy can be understood from the fact that the entropy has large weight on the largest Schmidt values,
but the required bond dimension is determined by the decay of Schmidt values in the tail.
Indeed, we show in Fig.~\ref{fig:Ising_Svst}(c) that the optimization of the entropy leads to a reduction in the first few Schmidt values accompanied by a slightly longer tail of small Schmidt values compared to $\Uanc=\id$.
Nevertheless, the tail decays faster than with backward time evolution.
In contrast, when the two-site disentangler described in Sec.~\ref{sec:normdisentangler} is used,
we can indeed slightly reduce the required maximal bond dimension as a proof of principle,
even though in practice performing the optimization itself is computationally more expensive than the speed-up gained by
the reduced bond dimension. In this case, the disentangler acts almost trivially in the region where $S^{+}_{L/2}$ was
applied, such that a larger tail of the singular values is avoided.
While this optimization reduced the bond dimension during the real-time evolution, in the case of imaginary-time
evolution we were not able to reduce the bond dimension with the same method.

\begin{figure}
	\includegraphics[width=\columnwidth]{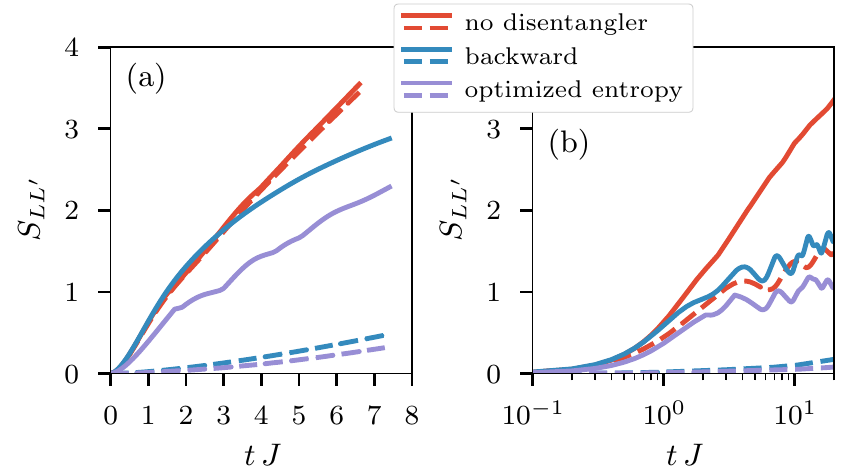}
	\caption{Comparison of the entanglement in the purification state $\ket{S^{+}_{L/2}(t,\beta=0)} = e^{-\im t H} S^{+}_{L/2}\ket{\psi_0}$
		for the Heisenberg chain \eqref{eq:Heisenberg} with $L=80$ sites without disorder ($W=0$) (a)
		and for a single disorder realization with $W=5J$ (b).
		In both panels, different colors compare different disentanglers $\Uanc$, 
		and solid lines (dashed lines) show the maximum (mean) over different bonds,
	}
	\label{fig:Heisenberg_Svst}
\end{figure}

As a second example, we consider the $S=1/2$ Heisenberg chain with disordered $z$-directed field,
\begin{equation}
	H = J \sum_{i=1}^{L-1}  \vec{S}_i \cdot \vec{S}_{i+1} - \sum_{i=1}^{L} h^z_i S^z_i,
\label{eq:Heisenberg}
\end{equation}
where $h^z_i$ is chosen uniformly in the interval $[-W, W]$.
This model has been established as a standard model in the study of many-body localization (MBL)  \cite{Basko2006,Gornyi2005,Abanin2017} in one dimension.
Numerically, a localization transition was found to occur at $W_c \approx 3.5 J$ \cite{Pal2010,Luitz2015}.
Figure~\ref{fig:Heisenberg_Svst}(a) again compares the entanglement growth of $\ket{S^{+}_{L/2}(t,\beta=0)}$ for the three choices of $\Uanc$ in the clean Heisenberg chain, $W=0$.
While the entropy grows linearly when no disentangler is used, 
the integrability of the Heisenberg chain and the presence of $S^z$ conservation restricts the entanglement of time-evolved local operators in the Heisenberg picture (here the ``backward'' evolution) to $S(t) \propto \log(t)$ \cite{Muth2011}.
Our results are compatible with the same $S(t) \propto \log(t)$ scaling when optimized, again with a smaller prefactor.
In the MBL phase [Fig.~\ref{fig:Heisenberg_Svst}(b)], even $\Uanc=\id$ displays only a logarithmic entanglement growth,
which is a characteristic feature of the MBL phase \cite{Bardarson2012,Znidaric2008,Vosk2013,Serbyn2013}.

\begin{figure}
    \includegraphics[width=\columnwidth]{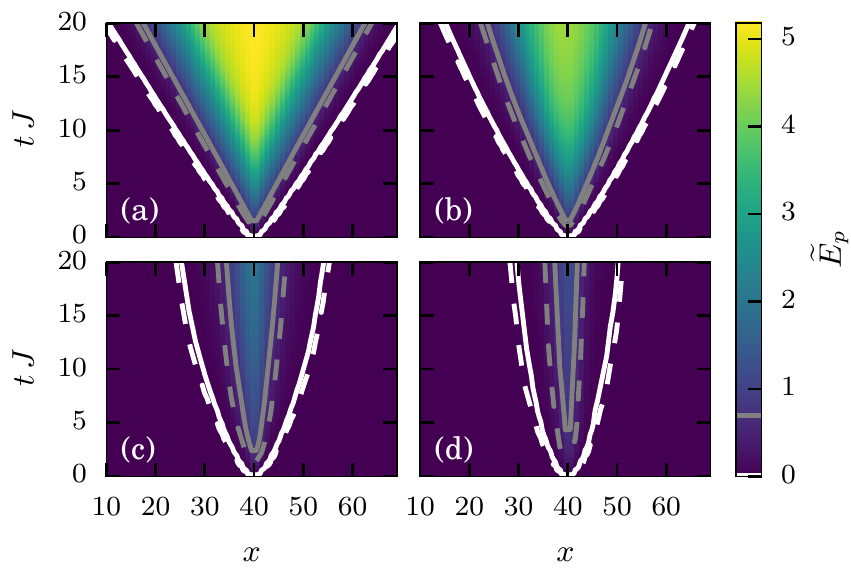}
    \caption{Optimized entanglement entropy $\entropyopt$ in the state $e^{-\im tH} S^{+}_{L/2}\ket{\psi_0}$ for the
		Heisenberg model \eqref{eq:Heisenberg} with disorder strength (a) $W=0$, (b) $W=J$, (c) $W=3J$, and (d) $W=5J$,  each averaged over 30 disorder realizations.
    The white and gray solid lines show the contour for the onset of finite values at a threshold of 0.01 and $\log(2)$.
    The dashed lines shows contours for the same threshold when backward time evolution is used.}
    \label{fig:MBL_S}
\end{figure}

Next, we focus on the spatial spread of the entanglement in $S^{+}_{L/2}\ket{\psi_0}$ when using the optimized disentangler,
tracking $\entropy$ as a function of time $t$, and bipartition bond $x$, shown in Fig.~\ref{fig:MBL_S}.
In the thermalizing regime, at small $W$ [Figs.~\ref{fig:MBL_S}(a) and (b)], we observe the expected linear light-cone spreading \cite{Kim2013}.
Deep in the MBL phase [Fig.~\ref{fig:MBL_S}(d)] we find a qualitatively different spreading which is compatible with a logarithmic light cone.
This is as expected from a generalized Lieb-Robinson bound $\mathbb{E}\| [A_i(t),B_j]\|\leq c\,t e^{\frac{|i-j|}{2\xi}} $, 
where $\xi$ is the localization length and $c >0$ some constant \cite{Kim2014a,Deng2016}.
At intermediate disorder, near the MBL transition, we observe a sub-linear spreading of the entanglement.
Although there are extended eigenstates in this region, the system is expected to be subdiffusive and exhibits only slow transport on very long time scales (inaccessible to our numerics) \cite{BarLev2015,Agarwal14,Chandran2016,Luitz2016,Luitz2016b,Luitz2017}.
Since the backward time evolution already reduces $\entropy$ to a zero (up to exponentially small corrections) outside of the light cone, it is not surprising that the contours of the onset are nearly unchanged compared to our optimized case.

\section{Conclusions}
We introduced an MPS based method to find a unitary $\Uanc$ acting on the ancilla DoF of a purification state, which reduces the entanglement both in equilibrium and during real-time evolution, at a similar cost to the TEBD algorithm.
At low temperatures, the optimized entanglement entropy $\entropyopt$ is half as large as in the TFD state, providing evidence that the algorithm actually finds the entanglement of purification $E_p$.
We find a maximum of $\entropyopt$ at intermediate $\beta$, the location of which diverges to $\betamax\rightarrow \infty$ as the gap closes.
During real-time evolution, the entanglement is significantly reduced both compared to $\Uanc=\id$ and backward time evolution.
In the clean Heisenberg chain, $\entropyopt$ shows a linear light-cone structure, which turns to a logarithmic spreading in the MBL phase (at large disorder).
The minimization of the entanglement is, however, not directly accompanied by a reduction of the required bond dimensions, as it leads to a larger tail of small Schmidt values.
This limitation might be overcome by another choice of local disentanglers.

\section*{Acknowledgments}
We acknowledge useful discussions with L.~Schoonderwoerd.
We have used the TeNPy package, which contains an implementation of our algorithm \cite{tenpy}.
This work was partially supported by the ERC Starting Grant No.~679722.
F.P.~acknowledges the support of the DFG Research Unit FOR 1807 through Grants No.~PO 1370/2-1 and No.~TRR80, the Nanosystems Initiative Munich (NIM) by the German Excellence Initiative, and the European Research Council (ERC) under the European Union's Horizon 2020 research and innovation program (grant agreement no.~771537).

\bibliography{projects-MEPS}

\begin{thebibliography}{55}%
\makeatletter
\providecommand \@ifxundefined [1]{%
 \@ifx{#1\undefined}
}%
\providecommand \@ifnum [1]{%
 \ifnum #1\expandafter \@firstoftwo
 \else \expandafter \@secondoftwo
 \fi
}%
\providecommand \@ifx [1]{%
 \ifx #1\expandafter \@firstoftwo
 \else \expandafter \@secondoftwo
 \fi
}%
\providecommand \natexlab [1]{#1}%
\providecommand \enquote  [1]{``#1''}%
\providecommand \bibnamefont  [1]{#1}%
\providecommand \bibfnamefont [1]{#1}%
\providecommand \citenamefont [1]{#1}%
\providecommand \href@noop [0]{\@secondoftwo}%
\providecommand \href [0]{\begingroup \@sanitize@url \@href}%
\providecommand \@href[1]{\@@startlink{#1}\@@href}%
\providecommand \@@href[1]{\endgroup#1\@@endlink}%
\providecommand \@sanitize@url [0]{\catcode `\\12\catcode `\$12\catcode
  `\&12\catcode `\#12\catcode `\^12\catcode `\_12\catcode `\%12\relax}%
\providecommand \@@startlink[1]{}%
\providecommand \@@endlink[0]{}%
\providecommand \url  [0]{\begingroup\@sanitize@url \@url }%
\providecommand \@url [1]{\endgroup\@href {#1}{\urlprefix }}%
\providecommand \urlprefix  [0]{URL }%
\providecommand \Eprint [0]{\href }%
\@ifxundefined \urlstyle {%
  \providecommand \doi  [0]{\begingroup \@sanitize@url \@doi}%
  \providecommand \@doi [1]{\endgroup \@@startlink {\doibase
  #1}doi:\discretionary {}{}{}#1\@@endlink }%
}{%
  \providecommand \doi  [0]{doi:\discretionary{}{}{}\begingroup
  \urlstyle{rm}\Url }%
}%
\providecommand \doibase [0]{http://dx.doi.org/}%
\providecommand \Doi [0]{\begingroup \@sanitize@url \@Doi }%
\providecommand \@Doi  [1]{\endgroup\@@startlink{\doibase#1}\@@Doi}%
\providecommand \@@Doi [1]{#1\@@endlink}%
\providecommand \selectlanguage [0]{\@gobble}%
\providecommand \bibinfo  [0]{\@secondoftwo}%
\providecommand \bibfield  [0]{\@secondoftwo}%
\providecommand \translation [1]{[#1]}%
\providecommand \BibitemOpen [0]{}%
\providecommand \bibitemStop [0]{}%
\providecommand \bibitemNoStop [0]{.\EOS\space}%
\providecommand \EOS [0]{\spacefactor3000\relax}%
\providecommand \BibitemShut  [1]{\csname bibitem#1\endcsname}%
\bibitem [{\citenamefont {Fannes}\ \emph {et~al.}(1992)\citenamefont {Fannes},
  \citenamefont {Nachtergaele},\ and\ \citenamefont {Werner}}]{Fannes-1992}%
  \BibitemOpen
  \bibfield  {author} {\bibinfo {author} {\bibfnamefont {M.}~\bibnamefont
  {Fannes}}, \bibinfo {author} {\bibfnamefont {B.}~\bibnamefont
  {Nachtergaele}}, \ and\ \bibinfo {author} {\bibfnamefont {R.~F.}\
  \bibnamefont {Werner}},\ }\Doi {10.1007/BF02099178} {\bibfield  {journal}
  {\bibinfo  {journal} {Commun. Math. Phys.},\ }\textbf {\bibinfo {volume}
  {144}},\ \bibinfo {pages} {443} (\bibinfo {year} {1992})},\ ISSN \bibinfo
  {issn} {00103616}\BibitemShut {NoStop}%
\bibitem [{\citenamefont {Hastings}(2006)}]{Hastings2006}%
  \BibitemOpen
  \bibfield  {author} {\bibinfo {author} {\bibfnamefont {M.~B.}\ \bibnamefont
  {Hastings}},\ }\Doi {10.1103/PhysRevB.73.085115} {\bibfield  {journal}
  {\bibinfo  {journal} {Phys. Rev. B},\ }\textbf {\bibinfo {volume} {73}},\
  \bibinfo {pages} {085115} (\bibinfo {year} {2006})},\ ISSN \bibinfo {issn}
  {1098-0121},\ \Eprint {http://arxiv.org/abs/0508554} {arXiv:0508554
  [cond-mat]} \BibitemShut {NoStop}%
\bibitem [{\citenamefont {Schollw{\"{o}}ck}(2011)}]{Schollwock2011}%
  \BibitemOpen
  \bibfield  {author} {\bibinfo {author} {\bibfnamefont {U.}~\bibnamefont
  {Schollw{\"{o}}ck}},\ }\Doi {10.1016/j.aop.2010.09.012} {\bibfield  {journal}
  {\bibinfo  {journal} {Ann. Phys. (N. Y).},\ }\textbf {\bibinfo {volume}
  {326}},\ \bibinfo {pages} {96} (\bibinfo {year} {2011})},\ ISSN \bibinfo
  {issn} {00034916},\ \Eprint {http://arxiv.org/abs/1008.3477}
  {arXiv:1008.3477} \BibitemShut {NoStop}%
\bibitem [{\citenamefont {Verstraete}\ and\ \citenamefont
  {Cirac}(2004)}]{Verstraete2004PEPS}%
  \BibitemOpen
  \bibfield  {author} {\bibinfo {author} {\bibfnamefont {F.}~\bibnamefont
  {Verstraete}}\ and\ \bibinfo {author} {\bibfnamefont {J.~I.}\ \bibnamefont
  {Cirac}},\ }\href {http://arxiv.org/abs/cond-mat/0407066} {\bibfield
  {journal} {\bibinfo  {journal} {arXiv:cond-mat/0407066}} (\bibinfo {year}
  {2004})},\ \Eprint {http://arxiv.org/abs/0407066} {arXiv:0407066 [cond-mat]}
  \BibitemShut {NoStop}%
\bibitem [{\citenamefont {White}(1992)}]{White1992DMRG}%
  \BibitemOpen
  \bibfield  {author} {\bibinfo {author} {\bibfnamefont {S.~R.}\ \bibnamefont
  {White}},\ }\Doi {10.1103/PhysRevLett.69.2863} {\bibfield  {journal}
  {\bibinfo  {journal} {Phys. Rev. Lett.},\ }\textbf {\bibinfo {volume} {69}},\
  \bibinfo {pages} {2863} (\bibinfo {year} {1992})},\ ISSN \bibinfo {issn}
  {0031-9007}\BibitemShut {NoStop}%
\bibitem [{\citenamefont {Vidal}(2004)}]{Vidal2004TEBD}%
  \BibitemOpen
  \bibfield  {author} {\bibinfo {author} {\bibfnamefont {G.}~\bibnamefont
  {Vidal}},\ }\Doi {10.1103/PhysRevLett.93.040502} {\bibfield  {journal}
  {\bibinfo  {journal} {Phys. Rev. Lett.},\ }\textbf {\bibinfo {volume} {93}},\
  \bibinfo {pages} {040502} (\bibinfo {year} {2004})},\ ISSN \bibinfo {issn}
  {00319007},\ \Eprint {http://arxiv.org/abs/0310089} {arXiv:0310089
  [quant-ph]} \BibitemShut {NoStop}%
\bibitem [{\citenamefont {Daley}\ \emph {et~al.}(2004)\citenamefont {Daley},
  \citenamefont {Kollath}, \citenamefont {Schollw{\"{o}}ck},\ and\
  \citenamefont {Vidal}}]{Daley2004tDMRG}%
  \BibitemOpen
  \bibfield  {author} {\bibinfo {author} {\bibfnamefont {A.~J.}\ \bibnamefont
  {Daley}}, \bibinfo {author} {\bibfnamefont {C.}~\bibnamefont {Kollath}},
  \bibinfo {author} {\bibfnamefont {U.}~\bibnamefont {Schollw{\"{o}}ck}}, \
  and\ \bibinfo {author} {\bibfnamefont {G.}~\bibnamefont {Vidal}},\ }\Doi
  {10.1088/1742-5468/2004/04/P04005} {\bibfield  {journal} {\bibinfo  {journal}
  {J. Stat. Mech. Theory Exp.},\ }\textbf {\bibinfo {volume} {2004}},\ \bibinfo
  {pages} {P04005} (\bibinfo {year} {2004})},\ ISSN \bibinfo {issn}
  {1742-5468},\ \Eprint {http://arxiv.org/abs/0403313} {arXiv:0403313
  [cond-mat]} \BibitemShut {NoStop}%
\bibitem [{\citenamefont {White}\ and\ \citenamefont
  {Feiguin}(2004)}]{White2004tDMRG}%
  \BibitemOpen
  \bibfield  {author} {\bibinfo {author} {\bibfnamefont {S.~R.}\ \bibnamefont
  {White}}\ and\ \bibinfo {author} {\bibfnamefont {A.~E.}\ \bibnamefont
  {Feiguin}},\ }\Doi {10.1103/PhysRevLett.93.076401} {\bibfield  {journal}
  {\bibinfo  {journal} {Phys. Rev. Lett.},\ }\textbf {\bibinfo {volume} {93}},\
  \bibinfo {pages} {076401} (\bibinfo {year} {2004})},\ ISSN \bibinfo {issn}
  {0031-9007},\ \Eprint {http://arxiv.org/abs/0403310} {arXiv:0403310
  [cond-mat]} \BibitemShut {NoStop}%
\bibitem [{\citenamefont {Barthel}(2017)}]{Barthel2017}%
  \BibitemOpen
  \bibfield  {author} {\bibinfo {author} {\bibfnamefont {T.}~\bibnamefont
  {Barthel}},\ }\href {http://arxiv.org/abs/1708.09349} {\bibfield  {journal}
  {\bibinfo  {journal} {arXiv:1708.09349}} (\bibinfo {year} {2017})},\ \Eprint
  {http://arxiv.org/abs/1708.09349} {arXiv:1708.09349} \BibitemShut {NoStop}%
\bibitem [{\citenamefont {Dubail}(2017)}]{Dubail2017}%
  \BibitemOpen
  \bibfield  {author} {\bibinfo {author} {\bibfnamefont {J.}~\bibnamefont
  {Dubail}},\ }\Doi {10.1088/1751-8121/aa6f38} {\bibfield  {journal} {\bibinfo
  {journal} {J. Phys. A Math. Theor.},\ }\textbf {\bibinfo {volume} {50}},\
  \bibinfo {pages} {234001} (\bibinfo {year} {2017})},\ ISSN \bibinfo {issn}
  {1751-8113},\ \Eprint {http://arxiv.org/abs/1612.08630} {arXiv:1612.08630}
  \BibitemShut {NoStop}%
\bibitem [{\citenamefont {Zwolak}\ and\ \citenamefont
  {Vidal}(2004)}]{Zwolak2004}%
  \BibitemOpen
  \bibfield  {author} {\bibinfo {author} {\bibfnamefont {M.}~\bibnamefont
  {Zwolak}}\ and\ \bibinfo {author} {\bibfnamefont {G.}~\bibnamefont {Vidal}},\
  }\Doi {10.1103/PhysRevLett.93.207205} {\bibfield  {journal} {\bibinfo
  {journal} {Phys. Rev. Lett.},\ }\textbf {\bibinfo {volume} {93}},\ \bibinfo
  {pages} {207205} (\bibinfo {year} {2004})},\ ISSN \bibinfo {issn}
  {0031-9007},\ \Eprint {http://arxiv.org/abs/0406440} {arXiv:0406440
  [cond-mat]} \BibitemShut {NoStop}%
\bibitem [{\citenamefont {White}(2009)}]{White2009}%
  \BibitemOpen
  \bibfield  {author} {\bibinfo {author} {\bibfnamefont {S.~R.}\ \bibnamefont
  {White}},\ }\Doi {10.1103/PhysRevLett.102.190601} {\bibfield  {journal}
  {\bibinfo  {journal} {Phys. Rev. Lett.},\ }\textbf {\bibinfo {volume}
  {102}},\ \bibinfo {pages} {190601} (\bibinfo {year} {2009})},\ ISSN \bibinfo
  {issn} {0031-9007},\ \Eprint {http://arxiv.org/abs/0902.4475}
  {arXiv:0902.4475} \BibitemShut {NoStop}%
\bibitem [{\citenamefont {Stoudenmire}\ and\ \citenamefont
  {White}(2010)}]{Stoudenmire2010}%
  \BibitemOpen
  \bibfield  {author} {\bibinfo {author} {\bibfnamefont {E.~M.}\ \bibnamefont
  {Stoudenmire}}\ and\ \bibinfo {author} {\bibfnamefont {S.~R.}\ \bibnamefont
  {White}},\ }\Doi {10.1088/1367-2630/12/5/055026} {\bibfield  {journal}
  {\bibinfo  {journal} {New J. Phys.},\ }\textbf {\bibinfo {volume} {12}},\
  \bibinfo {pages} {055026} (\bibinfo {year} {2010})},\ ISSN \bibinfo {issn}
  {13672630},\ \Eprint {http://arxiv.org/abs/1002.1305} {arXiv:1002.1305}
  \BibitemShut {NoStop}%
\bibitem [{\citenamefont {Binder}\ and\ \citenamefont
  {Barthel}(2017)}]{Binder2017}%
  \BibitemOpen
  \bibfield  {author} {\bibinfo {author} {\bibfnamefont {M.}~\bibnamefont
  {Binder}}\ and\ \bibinfo {author} {\bibfnamefont {T.}~\bibnamefont
  {Barthel}},\ }\Doi {10.1103/PhysRevB.95.195148} {\bibfield  {journal}
  {\bibinfo  {journal} {Phys. Rev. B},\ }\textbf {\bibinfo {volume} {95}},\
  \bibinfo {pages} {195148} (\bibinfo {year} {2017})},\ ISSN \bibinfo {issn}
  {2469-9950},\ \Eprint {http://arxiv.org/abs/1701.03872} {arXiv:1701.03872}
  \BibitemShut {NoStop}%
\bibitem [{\citenamefont {Berta}\ \emph {et~al.}(2017)\citenamefont {Berta},
  \citenamefont {Brandao}, \citenamefont {Haegeman}, \citenamefont {Scholz},\
  and\ \citenamefont {Verstraete}}]{Berta2017}%
  \BibitemOpen
  \bibfield  {author} {\bibinfo {author} {\bibfnamefont {M.}~\bibnamefont
  {Berta}}, \bibinfo {author} {\bibfnamefont {F.~G. S.~L.}\ \bibnamefont
  {Brandao}}, \bibinfo {author} {\bibfnamefont {J.}~\bibnamefont {Haegeman}},
  \bibinfo {author} {\bibfnamefont {V.~B.}\ \bibnamefont {Scholz}}, \ and\
  \bibinfo {author} {\bibfnamefont {F.}~\bibnamefont {Verstraete}},\ }\href
  {http://arxiv.org/abs/1709.07423} {\bibfield  {journal} {\bibinfo  {journal}
  {arXiv:1709.07423}} (\bibinfo {year} {2017})},\ \Eprint
  {http://arxiv.org/abs/1709.07423} {arXiv:1709.07423} \BibitemShut {NoStop}%
\bibitem [{\citenamefont {Verstraete}\ \emph {et~al.}(2004)\citenamefont
  {Verstraete}, \citenamefont {Garc{\'{i}}a-Ripoll},\ and\ \citenamefont
  {Cirac}}]{Verstraete2004Purification}%
  \BibitemOpen
  \bibfield  {author} {\bibinfo {author} {\bibfnamefont {F.}~\bibnamefont
  {Verstraete}}, \bibinfo {author} {\bibfnamefont {J.~J.}\ \bibnamefont
  {Garc{\'{i}}a-Ripoll}}, \ and\ \bibinfo {author} {\bibfnamefont {J.~I.}\
  \bibnamefont {Cirac}},\ }\Doi {10.1103/PhysRevLett.93.207204} {\bibfield
  {journal} {\bibinfo  {journal} {Phys. Rev. Lett.},\ }\textbf {\bibinfo
  {volume} {93}},\ \bibinfo {pages} {207204} (\bibinfo {year} {2004})},\ ISSN
  \bibinfo {issn} {0031-9007},\ \Eprint {http://arxiv.org/abs/0406426}
  {arXiv:0406426 [cond-mat]} \BibitemShut {NoStop}%
\bibitem [{\citenamefont {Barthel}\ \emph {et~al.}(2009)\citenamefont
  {Barthel}, \citenamefont {Schollw{\"{o}}ck},\ and\ \citenamefont
  {White}}]{Barthel2009}%
  \BibitemOpen
  \bibfield  {author} {\bibinfo {author} {\bibfnamefont {T.}~\bibnamefont
  {Barthel}}, \bibinfo {author} {\bibfnamefont {U.}~\bibnamefont
  {Schollw{\"{o}}ck}}, \ and\ \bibinfo {author} {\bibfnamefont {S.~R.}\
  \bibnamefont {White}},\ }\Doi {10.1103/PhysRevB.79.245101} {\bibfield
  {journal} {\bibinfo  {journal} {Phys. Rev. B},\ }\textbf {\bibinfo {volume}
  {79}},\ \bibinfo {pages} {245101} (\bibinfo {year} {2009})},\ ISSN \bibinfo
  {issn} {1098-0121},\ \Eprint {http://arxiv.org/abs/0901.2342}
  {arXiv:0901.2342} \BibitemShut {NoStop}%
\bibitem [{\citenamefont {{De las Cuevas}}\ \emph {et~al.}(2013)\citenamefont
  {{De las Cuevas}}, \citenamefont {Schuch}, \citenamefont
  {P{\'{e}}rez-Garc{\'{i}}a},\ and\ \citenamefont {{Ignacio
  Cirac}}}]{deLasCuevas2013}%
  \BibitemOpen
  \bibfield  {author} {\bibinfo {author} {\bibfnamefont {G.}~\bibnamefont {{De
  las Cuevas}}}, \bibinfo {author} {\bibfnamefont {N.}~\bibnamefont {Schuch}},
  \bibinfo {author} {\bibfnamefont {D.}~\bibnamefont
  {P{\'{e}}rez-Garc{\'{i}}a}}, \ and\ \bibinfo {author} {\bibfnamefont
  {J.}~\bibnamefont {{Ignacio Cirac}}},\ }\Doi {10.1088/1367-2630/15/12/123021}
  {\bibfield  {journal} {\bibinfo  {journal} {New J. Phys.},\ }\textbf
  {\bibinfo {volume} {15}},\ \bibinfo {pages} {123021} (\bibinfo {year}
  {2013})},\ ISSN \bibinfo {issn} {1367-2630},\ \Eprint
  {http://arxiv.org/abs/1308.1914} {arXiv:1308.1914} \BibitemShut {NoStop}%
\bibitem [{\citenamefont {{De las Cuevas}}\ \emph {et~al.}(2016)\citenamefont
  {{De las Cuevas}}, \citenamefont {Cubitt}, \citenamefont {Cirac},
  \citenamefont {Wolf},\ and\ \citenamefont
  {P{\'{e}}rez-Garc{\'{i}}a}}]{deLasCuevas2016}%
  \BibitemOpen
  \bibfield  {author} {\bibinfo {author} {\bibfnamefont {G.}~\bibnamefont {{De
  las Cuevas}}}, \bibinfo {author} {\bibfnamefont {T.~S.}\ \bibnamefont
  {Cubitt}}, \bibinfo {author} {\bibfnamefont {J.~I.}\ \bibnamefont {Cirac}},
  \bibinfo {author} {\bibfnamefont {M.~M.}\ \bibnamefont {Wolf}}, \ and\
  \bibinfo {author} {\bibfnamefont {D.}~\bibnamefont
  {P{\'{e}}rez-Garc{\'{i}}a}},\ }\Doi {10.1063/1.4954983} {\bibfield  {journal}
  {\bibinfo  {journal} {J. Math. Phys.},\ }\textbf {\bibinfo {volume} {57}},\
  \bibinfo {pages} {071902} (\bibinfo {year} {2016})},\ ISSN \bibinfo {issn}
  {0022-2488},\ \Eprint {http://arxiv.org/abs/1512.05709} {arXiv:1512.05709}
  \BibitemShut {NoStop}%
\bibitem [{\citenamefont {Karrasch}\ \emph {et~al.}(2012)\citenamefont
  {Karrasch}, \citenamefont {Bardarson},\ and\ \citenamefont
  {Moore}}]{Karrasch2012}%
  \BibitemOpen
  \bibfield  {author} {\bibinfo {author} {\bibfnamefont {C.}~\bibnamefont
  {Karrasch}}, \bibinfo {author} {\bibfnamefont {J.~H.}\ \bibnamefont
  {Bardarson}}, \ and\ \bibinfo {author} {\bibfnamefont {J.~E.}\ \bibnamefont
  {Moore}},\ }\Doi {10.1103/PhysRevLett.108.227206} {\bibfield  {journal}
  {\bibinfo  {journal} {Phys. Rev. Lett.},\ }\textbf {\bibinfo {volume}
  {108}},\ \bibinfo {pages} {227206} (\bibinfo {year} {2012})},\ ISSN \bibinfo
  {issn} {00319007},\ \Eprint {http://arxiv.org/abs/1111.4508}
  {arXiv:1111.4508} \BibitemShut {NoStop}%
\bibitem [{\citenamefont {Barthel}(2013)}]{Barthel2013}%
  \BibitemOpen
  \bibfield  {author} {\bibinfo {author} {\bibfnamefont {T.}~\bibnamefont
  {Barthel}},\ }\Doi {10.1088/1367-2630/15/7/073010} {\bibfield  {journal}
  {\bibinfo  {journal} {New J. Phys.},\ }\textbf {\bibinfo {volume} {15}},\
  \bibinfo {pages} {073010} (\bibinfo {year} {2013})},\ ISSN \bibinfo {issn}
  {1367-2630},\ \Eprint {http://arxiv.org/abs/1301.2246} {arXiv:1301.2246}
  \BibitemShut {NoStop}%
\bibitem [{\citenamefont {Terhal}\ \emph {et~al.}(2002)\citenamefont {Terhal},
  \citenamefont {Horodecki}, \citenamefont {Leung},\ and\ \citenamefont
  {DiVincenzo}}]{Terhal2002}%
  \BibitemOpen
  \bibfield  {author} {\bibinfo {author} {\bibfnamefont {B.~M.}\ \bibnamefont
  {Terhal}}, \bibinfo {author} {\bibfnamefont {M.}~\bibnamefont {Horodecki}},
  \bibinfo {author} {\bibfnamefont {D.~W.}\ \bibnamefont {Leung}}, \ and\
  \bibinfo {author} {\bibfnamefont {D.~P.}\ \bibnamefont {DiVincenzo}},\ }\Doi
  {10.1063/1.1498001} {\bibfield  {journal} {\bibinfo  {journal} {J. Math.
  Phys.},\ }\textbf {\bibinfo {volume} {43}},\ \bibinfo {pages} {4286}
  (\bibinfo {year} {2002})},\ ISSN \bibinfo {issn} {0022-2488},\ \Eprint
  {http://arxiv.org/abs/0202044} {arXiv:0202044 [quant-ph]} \BibitemShut
  {NoStop}%
\bibitem [{\citenamefont {Schuch}\ \emph {et~al.}(2008)\citenamefont {Schuch},
  \citenamefont {Wolf}, \citenamefont {Verstraete},\ and\ \citenamefont
  {Cirac}}]{Schuch2008}%
  \BibitemOpen
  \bibfield  {author} {\bibinfo {author} {\bibfnamefont {N.}~\bibnamefont
  {Schuch}}, \bibinfo {author} {\bibfnamefont {M.~M.}\ \bibnamefont {Wolf}},
  \bibinfo {author} {\bibfnamefont {F.}~\bibnamefont {Verstraete}}, \ and\
  \bibinfo {author} {\bibfnamefont {J.~I.}\ \bibnamefont {Cirac}},\ }\Doi
  {10.1103/PhysRevLett.100.030504} {\bibfield  {journal} {\bibinfo  {journal}
  {Phys. Rev. Lett.},\ }\textbf {\bibinfo {volume} {100}},\ \bibinfo {pages}
  {030504} (\bibinfo {year} {2008})},\ ISSN \bibinfo {issn} {0031-9007},\
  \Eprint {http://arxiv.org/abs/0705.0292} {arXiv:0705.0292} \BibitemShut
  {NoStop}%
\bibitem [{\citenamefont {Leviatan}\ \emph {et~al.}(2017)\citenamefont
  {Leviatan}, \citenamefont {Pollmann}, \citenamefont {Bardarson},\ and\
  \citenamefont {Altman}}]{Leviatan2017}%
  \BibitemOpen
  \bibfield  {author} {\bibinfo {author} {\bibfnamefont {E.}~\bibnamefont
  {Leviatan}}, \bibinfo {author} {\bibfnamefont {F.}~\bibnamefont {Pollmann}},
  \bibinfo {author} {\bibfnamefont {J.~H.}\ \bibnamefont {Bardarson}}, \ and\
  \bibinfo {author} {\bibfnamefont {E.}~\bibnamefont {Altman}},\ }\href
  {http://arxiv.org/abs/1702.08894} {\bibfield  {journal} {\bibinfo  {journal}
  {arXiv:1702.08894}} (\bibinfo {year} {2017})},\ \Eprint
  {http://arxiv.org/abs/1702.08894} {arXiv:1702.08894} \BibitemShut {NoStop}%
\bibitem [{\citenamefont {White}\ \emph {et~al.}(2018)\citenamefont {White},
  \citenamefont {Zaletel}, \citenamefont {Mong},\ and\ \citenamefont
  {Refael}}]{White2017}%
  \BibitemOpen
  \bibfield  {author} {\bibinfo {author} {\bibfnamefont {C.~D.}\ \bibnamefont
  {White}}, \bibinfo {author} {\bibfnamefont {M.}~\bibnamefont {Zaletel}},
  \bibinfo {author} {\bibfnamefont {R.~S.~K.}\ \bibnamefont {Mong}}, \ and\
  \bibinfo {author} {\bibfnamefont {G.}~\bibnamefont {Refael}},\ }\Doi
  {10.1103/PhysRevB.97.035127} {\bibfield  {journal} {\bibinfo  {journal}
  {Phys. Rev. B},\ }\textbf {\bibinfo {volume} {97}},\ \bibinfo {pages}
  {035127} (\bibinfo {year} {2018})},\ ISSN \bibinfo {issn} {2469-9950},\
  \Eprint {http://arxiv.org/abs/1707.01506} {arXiv:1707.01506} \BibitemShut
  {NoStop}%
\bibitem [{\citenamefont {Haegeman}\ \emph {et~al.}(2011)\citenamefont
  {Haegeman}, \citenamefont {Cirac}, \citenamefont {Osborne}, \citenamefont
  {Pi{\v{z}}orn}, \citenamefont {Verschelde},\ and\ \citenamefont
  {Verstraete}}]{Haegeman2011}%
  \BibitemOpen
  \bibfield  {author} {\bibinfo {author} {\bibfnamefont {J.}~\bibnamefont
  {Haegeman}}, \bibinfo {author} {\bibfnamefont {J.~I.}\ \bibnamefont {Cirac}},
  \bibinfo {author} {\bibfnamefont {T.~J.}\ \bibnamefont {Osborne}}, \bibinfo
  {author} {\bibfnamefont {I.}~\bibnamefont {Pi{\v{z}}orn}}, \bibinfo {author}
  {\bibfnamefont {H.}~\bibnamefont {Verschelde}}, \ and\ \bibinfo {author}
  {\bibfnamefont {F.}~\bibnamefont {Verstraete}},\ }\Doi
  {10.1103/PhysRevLett.107.070601} {\bibfield  {journal} {\bibinfo  {journal}
  {Phys. Rev. Lett.},\ }\textbf {\bibinfo {volume} {107}},\ \bibinfo {pages}
  {070601} (\bibinfo {year} {2011})},\ ISSN \bibinfo {issn} {0031-9007},\
  \Eprint {http://arxiv.org/abs/1103.0936} {arXiv:1103.0936} \BibitemShut
  {NoStop}%
\bibitem [{\citenamefont {Zaletel}\ \emph {et~al.}(2015)\citenamefont
  {Zaletel}, \citenamefont {Mong}, \citenamefont {Karrasch}, \citenamefont
  {Moore},\ and\ \citenamefont {Pollmann}}]{Zaletel2015}%
  \BibitemOpen
  \bibfield  {author} {\bibinfo {author} {\bibfnamefont {M.~P.}\ \bibnamefont
  {Zaletel}}, \bibinfo {author} {\bibfnamefont {R.~S.~K.}\ \bibnamefont
  {Mong}}, \bibinfo {author} {\bibfnamefont {C.}~\bibnamefont {Karrasch}},
  \bibinfo {author} {\bibfnamefont {J.~E.}\ \bibnamefont {Moore}}, \ and\
  \bibinfo {author} {\bibfnamefont {F.}~\bibnamefont {Pollmann}},\ }\Doi
  {10.1103/PhysRevB.91.165112} {\bibfield  {journal} {\bibinfo  {journal}
  {Phys. Rev. B},\ }\textbf {\bibinfo {volume} {91}},\ \bibinfo {pages}
  {165112} (\bibinfo {year} {2015})},\ ISSN \bibinfo {issn} {1098-0121},\
  \Eprint {http://arxiv.org/abs/1407.1832} {arXiv:1407.1832} \BibitemShut
  {NoStop}%
\bibitem [{\citenamefont {Karrasch}\ \emph {et~al.}(2013)\citenamefont
  {Karrasch}, \citenamefont {Bardarson},\ and\ \citenamefont
  {Moore}}]{Karrasch2013}%
  \BibitemOpen
  \bibfield  {author} {\bibinfo {author} {\bibfnamefont {C.}~\bibnamefont
  {Karrasch}}, \bibinfo {author} {\bibfnamefont {J.~H.}\ \bibnamefont
  {Bardarson}}, \ and\ \bibinfo {author} {\bibfnamefont {J.~E.}\ \bibnamefont
  {Moore}},\ }\Doi {10.1088/1367-2630/15/8/083031} {\bibfield  {journal}
  {\bibinfo  {journal} {New J. Phys.},\ }\textbf {\bibinfo {volume} {15}},\
  \bibinfo {pages} {083031} (\bibinfo {year} {2013})},\ ISSN \bibinfo {issn}
  {1367-2630},\ \Eprint {http://arxiv.org/abs/1303.3942} {arXiv:1303.3942}
  \BibitemShut {NoStop}%
\bibitem [{\citenamefont {Evenbly}\ and\ \citenamefont
  {Vidal}(2014)}]{Evenbly2014}%
  \BibitemOpen
  \bibfield  {author} {\bibinfo {author} {\bibfnamefont {G.}~\bibnamefont
  {Evenbly}}\ and\ \bibinfo {author} {\bibfnamefont {G.}~\bibnamefont
  {Vidal}},\ }\Doi {10.1007/s10955-014-0983-1} {\bibfield  {journal} {\bibinfo
  {journal} {J. Stat. Phys.},\ }\textbf {\bibinfo {volume} {157}},\ \bibinfo
  {pages} {931} (\bibinfo {year} {2014})},\ ISSN \bibinfo {issn} {0022-4715},\
  \Eprint {http://arxiv.org/abs/1312.0303} {arXiv:1312.0303} \BibitemShut
  {NoStop}%
\bibitem [{\citenamefont {Nielsen}\ and\ \citenamefont
  {Chuang}(2002)}]{Nielsen2000}%
  \BibitemOpen
  \bibfield  {author} {\bibinfo {author} {\bibfnamefont {M.~A.}\ \bibnamefont
  {Nielsen}}\ and\ \bibinfo {author} {\bibfnamefont {I.~L.}\ \bibnamefont
  {Chuang}},\ }\Doi {10.1119/1.1463744} {\emph {\bibinfo {title} {Quantum
  Computation and Quantum Information}}}\ (\bibinfo  {publisher} {Cambridge
  University Press},\ \bibinfo {year} {2002})\ ISBN \bibinfo {isbn}
  {9781107002173}\BibitemShut {NoStop}%
\bibitem [{\citenamefont {Singh}\ \emph {et~al.}(2010)\citenamefont {Singh},
  \citenamefont {Pfeifer},\ and\ \citenamefont {Vidal}}]{Singh2010}%
  \BibitemOpen
  \bibfield  {author} {\bibinfo {author} {\bibfnamefont {S.}~\bibnamefont
  {Singh}}, \bibinfo {author} {\bibfnamefont {R.~N.~C.}\ \bibnamefont
  {Pfeifer}}, \ and\ \bibinfo {author} {\bibfnamefont {G.}~\bibnamefont
  {Vidal}},\ }\Doi {10.1103/PhysRevA.82.050301} {\bibfield  {journal} {\bibinfo
   {journal} {Phys. Rev. A},\ }\textbf {\bibinfo {volume} {82}},\ \bibinfo
  {pages} {050301} (\bibinfo {year} {2010})},\ ISSN \bibinfo {issn}
  {10502947},\ \Eprint {http://arxiv.org/abs/0907.2994} {arXiv:0907.2994}
  \BibitemShut {NoStop}%
\bibitem [{\citenamefont {Hauschild}\ and\ \citenamefont
  {Pollmann}(2018)}]{tenpy}%
  \BibitemOpen
  \bibfield  {author} {\bibinfo {author} {\bibfnamefont {J.}~\bibnamefont
  {Hauschild}}\ and\ \bibinfo {author} {\bibfnamefont {F.}~\bibnamefont
  {Pollmann}},\ }\Doi {10.21468/SciPostPhysLectNotes.5} {\bibfield  {journal}
  {\bibinfo  {journal} {SciPost Phys. Lect. Notes},\ \bibinfo {pages} {5}}
  (\bibinfo {year} {2018})},\ \bibinfo {note} {the code is available online at
  \url{https://github.com/tenpy/tenpy/}},\ \Eprint
  {http://arxiv.org/abs/1805.00055} {arXiv:1805.00055} \BibitemShut {NoStop}%
\bibitem [{\citenamefont {Harada}(2018)}]{Harada2018}%
  \BibitemOpen
  \bibfield  {author} {\bibinfo {author} {\bibfnamefont {K.}~\bibnamefont
  {Harada}},\ }\Doi {10.1103/PhysRevB.97.045124} {\bibfield  {journal}
  {\bibinfo  {journal} {Phys. Rev. B},\ }\textbf {\bibinfo {volume} {97}},\
  \bibinfo {pages} {045124} (\bibinfo {year} {2018})},\ ISSN \bibinfo {issn}
  {2469-9950},\ \Eprint {http://arxiv.org/abs/1710.01830} {arXiv:1710.01830}
  \BibitemShut {NoStop}%
\bibitem [{\citenamefont {Hyatt}\ \emph {et~al.}(2017)\citenamefont {Hyatt},
  \citenamefont {Garrison},\ and\ \citenamefont {Bauer}}]{Hyatt2017}%
  \BibitemOpen
  \bibfield  {author} {\bibinfo {author} {\bibfnamefont {K.}~\bibnamefont
  {Hyatt}}, \bibinfo {author} {\bibfnamefont {J.~R.}\ \bibnamefont {Garrison}},
  \ and\ \bibinfo {author} {\bibfnamefont {B.}~\bibnamefont {Bauer}},\ }\Doi
  {10.1103/PhysRevLett.119.140502} {\bibfield  {journal} {\bibinfo  {journal}
  {Phys. Rev. Lett.},\ }\textbf {\bibinfo {volume} {119}},\ \bibinfo {pages}
  {140502} (\bibinfo {year} {2017})},\ ISSN \bibinfo {issn} {0031-9007},\
  \Eprint {http://arxiv.org/abs/1704.01974} {arXiv:1704.01974} \BibitemShut
  {NoStop}%
\bibitem [{\citenamefont {Umemoto}\ and\ \citenamefont
  {Takayanagi}(2018)}]{Takayanagi2017}%
  \BibitemOpen
  \bibfield  {author} {\bibinfo {author} {\bibfnamefont {K.}~\bibnamefont
  {Umemoto}}\ and\ \bibinfo {author} {\bibfnamefont {T.}~\bibnamefont
  {Takayanagi}},\ }\Doi {10.1038/s41567-018-0075-2} {\bibfield  {journal}
  {\bibinfo  {journal} {Nat. Phys.},\ }\textbf {\bibinfo {volume} {14}},\
  \bibinfo {pages} {573} (\bibinfo {year} {2018})},\ ISSN \bibinfo {issn}
  {1745-2473},\ \Eprint {http://arxiv.org/abs/1708.09393} {arXiv:1708.09393}
  \BibitemShut {NoStop}%
\bibitem [{\citenamefont {Nguyen}\ \emph {et~al.}(2018)\citenamefont {Nguyen},
  \citenamefont {Devakul}, \citenamefont {Halbasch}, \citenamefont {Zaletel},\
  and\ \citenamefont {Swingle}}]{Nguyen2017}%
  \BibitemOpen
  \bibfield  {author} {\bibinfo {author} {\bibfnamefont {P.}~\bibnamefont
  {Nguyen}}, \bibinfo {author} {\bibfnamefont {T.}~\bibnamefont {Devakul}},
  \bibinfo {author} {\bibfnamefont {M.~G.}\ \bibnamefont {Halbasch}}, \bibinfo
  {author} {\bibfnamefont {M.~P.}\ \bibnamefont {Zaletel}}, \ and\ \bibinfo
  {author} {\bibfnamefont {B.}~\bibnamefont {Swingle}},\ }\Doi
  {10.1007/JHEP01(2018)098} {\bibfield  {journal} {\bibinfo  {journal} {J. High
  Energy Phys.},\ }\textbf {\bibinfo {volume} {2018}},\ \bibinfo {pages} {98}
  (\bibinfo {year} {2018})},\ ISSN \bibinfo {issn} {1029-8479},\ \Eprint
  {http://arxiv.org/abs/1709.07424} {arXiv:1709.07424} \BibitemShut {NoStop}%
\bibitem [{\citenamefont {Basko}\ \emph {et~al.}(2006)\citenamefont {Basko},
  \citenamefont {Aleiner},\ and\ \citenamefont {Altshuler}}]{Basko2006}%
  \BibitemOpen
  \bibfield  {author} {\bibinfo {author} {\bibfnamefont {D.~M.}\ \bibnamefont
  {Basko}}, \bibinfo {author} {\bibfnamefont {I.~L.}\ \bibnamefont {Aleiner}},
  \ and\ \bibinfo {author} {\bibfnamefont {B.~L.}\ \bibnamefont {Altshuler}},\
  }\Doi {10.1016/j.aop.2005.11.014} {\bibfield  {journal} {\bibinfo  {journal}
  {Ann. Phys. (N. Y).},\ }\textbf {\bibinfo {volume} {321}},\ \bibinfo {pages}
  {1126} (\bibinfo {year} {2006})},\ ISSN \bibinfo {issn} {00034916},\ \Eprint
  {http://arxiv.org/abs/0506617} {arXiv:0506617 [cond-mat]} \BibitemShut
  {NoStop}%
\bibitem [{\citenamefont {Gornyi}\ \emph {et~al.}(2005)\citenamefont {Gornyi},
  \citenamefont {Mirlin},\ and\ \citenamefont {Polyakov}}]{Gornyi2005}%
  \BibitemOpen
  \bibfield  {author} {\bibinfo {author} {\bibfnamefont {I.~V.}\ \bibnamefont
  {Gornyi}}, \bibinfo {author} {\bibfnamefont {A.~D.}\ \bibnamefont {Mirlin}},
  \ and\ \bibinfo {author} {\bibfnamefont {D.~G.}\ \bibnamefont {Polyakov}},\
  }\Doi {10.1103/PhysRevLett.95.206603} {\bibfield  {journal} {\bibinfo
  {journal} {Phys. Rev. Lett.},\ }\textbf {\bibinfo {volume} {95}},\ \bibinfo
  {pages} {206603} (\bibinfo {year} {2005})},\ ISSN \bibinfo {issn}
  {00319007},\ \Eprint {http://arxiv.org/abs/0506411} {arXiv:0506411
  [cond-mat]} \BibitemShut {NoStop}%
\bibitem [{\citenamefont {Abanin}\ and\ \citenamefont
  {Papi{\'{c}}}(2017)}]{Abanin2017}%
  \BibitemOpen
  \bibfield  {author} {\bibinfo {author} {\bibfnamefont {D.~A.}\ \bibnamefont
  {Abanin}}\ and\ \bibinfo {author} {\bibfnamefont {Z.}~\bibnamefont
  {Papi{\'{c}}}},\ }\Doi {10.1002/andp.201700169} {\bibfield  {journal}
  {\bibinfo  {journal} {Ann. Phys.},\ }\textbf {\bibinfo {volume} {529}},\
  \bibinfo {pages} {1700169} (\bibinfo {year} {2017})},\ ISSN \bibinfo {issn}
  {00033804},\ \Eprint {http://arxiv.org/abs/1705.09103} {arXiv:1705.09103}
  \BibitemShut {NoStop}%
\bibitem [{\citenamefont {Pal}\ and\ \citenamefont {Huse}(2010)}]{Pal2010}%
  \BibitemOpen
  \bibfield  {author} {\bibinfo {author} {\bibfnamefont {A.}~\bibnamefont
  {Pal}}\ and\ \bibinfo {author} {\bibfnamefont {D.~A.}\ \bibnamefont {Huse}},\
  }\Doi {10.1103/PhysRevB.82.174411} {\bibfield  {journal} {\bibinfo  {journal}
  {Phys. Rev. B},\ }\textbf {\bibinfo {volume} {82}},\ \bibinfo {pages}
  {174411} (\bibinfo {year} {2010})},\ ISSN \bibinfo {issn} {1098-0121},\
  \Eprint {http://arxiv.org/abs/1010.1992} {arXiv:1010.1992} \BibitemShut
  {NoStop}%
\bibitem [{\citenamefont {Luitz}\ \emph {et~al.}(2015)\citenamefont {Luitz},
  \citenamefont {Laflorencie},\ and\ \citenamefont {Alet}}]{Luitz2015}%
  \BibitemOpen
  \bibfield  {author} {\bibinfo {author} {\bibfnamefont {D.~J.}\ \bibnamefont
  {Luitz}}, \bibinfo {author} {\bibfnamefont {N.}~\bibnamefont {Laflorencie}},
  \ and\ \bibinfo {author} {\bibfnamefont {F.}~\bibnamefont {Alet}},\ }\Doi
  {10.1103/PhysRevB.91.081103} {\bibfield  {journal} {\bibinfo  {journal}
  {Phys. Rev. B},\ }\textbf {\bibinfo {volume} {91}},\ \bibinfo {pages}
  {081103} (\bibinfo {year} {2015})},\ ISSN \bibinfo {issn} {1098-0121},\
  \Eprint {http://arxiv.org/abs/1411.0660} {arXiv:1411.0660} \BibitemShut
  {NoStop}%
\bibitem [{\citenamefont {Muth}\ \emph {et~al.}(2011)\citenamefont {Muth},
  \citenamefont {Unanyan},\ and\ \citenamefont {Fleischhauer}}]{Muth2011}%
  \BibitemOpen
  \bibfield  {author} {\bibinfo {author} {\bibfnamefont {D.}~\bibnamefont
  {Muth}}, \bibinfo {author} {\bibfnamefont {R.~G.}\ \bibnamefont {Unanyan}}, \
  and\ \bibinfo {author} {\bibfnamefont {M.}~\bibnamefont {Fleischhauer}},\
  }\Doi {10.1103/PhysRevLett.106.077202} {\bibfield  {journal} {\bibinfo
  {journal} {Phys. Rev. Lett.},\ }\textbf {\bibinfo {volume} {106}},\ \bibinfo
  {pages} {077202} (\bibinfo {year} {2011})},\ ISSN \bibinfo {issn}
  {0031-9007},\ \Eprint {http://arxiv.org/abs/1009.4646} {arXiv:1009.4646}
  \BibitemShut {NoStop}%
\bibitem [{\citenamefont {Bardarson}\ \emph {et~al.}(2012)\citenamefont
  {Bardarson}, \citenamefont {Pollmann},\ and\ \citenamefont
  {Moore}}]{Bardarson2012}%
  \BibitemOpen
  \bibfield  {author} {\bibinfo {author} {\bibfnamefont {J.~H.}\ \bibnamefont
  {Bardarson}}, \bibinfo {author} {\bibfnamefont {F.}~\bibnamefont {Pollmann}},
  \ and\ \bibinfo {author} {\bibfnamefont {J.~E.}\ \bibnamefont {Moore}},\
  }\Doi {10.1103/PhysRevLett.109.017202} {\bibfield  {journal} {\bibinfo
  {journal} {Phys. Rev. Lett.},\ }\textbf {\bibinfo {volume} {109}},\ \bibinfo
  {pages} {017202} (\bibinfo {year} {2012})},\ ISSN \bibinfo {issn}
  {0031-9007},\ \Eprint {http://arxiv.org/abs/1202.5532} {arXiv:1202.5532}
  \BibitemShut {NoStop}%
\bibitem [{\citenamefont {{\v{Z}}nidari{\v{c}}}\ \emph
  {et~al.}(2008)\citenamefont {{\v{Z}}nidari{\v{c}}}, \citenamefont {Prosen},\
  and\ \citenamefont {Prelov{\v{s}}ek}}]{Znidaric2008}%
  \BibitemOpen
  \bibfield  {author} {\bibinfo {author} {\bibfnamefont {M.}~\bibnamefont
  {{\v{Z}}nidari{\v{c}}}}, \bibinfo {author} {\bibfnamefont {T.}~\bibnamefont
  {Prosen}}, \ and\ \bibinfo {author} {\bibfnamefont {P.}~\bibnamefont
  {Prelov{\v{s}}ek}},\ }\Doi {10.1103/PhysRevB.77.064426} {\bibfield  {journal}
  {\bibinfo  {journal} {Phys. Rev. B},\ }\textbf {\bibinfo {volume} {77}},\
  \bibinfo {pages} {064426} (\bibinfo {year} {2008})},\ ISSN \bibinfo {issn}
  {1098-0121},\ \Eprint {http://arxiv.org/abs/0706.2539} {arXiv:0706.2539}
  \BibitemShut {NoStop}%
\bibitem [{\citenamefont {Vosk}\ and\ \citenamefont {Altman}(2013)}]{Vosk2013}%
  \BibitemOpen
  \bibfield  {author} {\bibinfo {author} {\bibfnamefont {R.}~\bibnamefont
  {Vosk}}\ and\ \bibinfo {author} {\bibfnamefont {E.}~\bibnamefont {Altman}},\
  }\Doi {10.1103/PhysRevLett.110.067204} {\bibfield  {journal} {\bibinfo
  {journal} {Phys. Rev. Lett.},\ }\textbf {\bibinfo {volume} {110}},\ \bibinfo
  {pages} {067204} (\bibinfo {year} {2013})},\ ISSN \bibinfo {issn}
  {0031-9007},\ \Eprint {http://arxiv.org/abs/1205.0026} {arXiv:1205.0026}
  \BibitemShut {NoStop}%
\bibitem [{\citenamefont {Serbyn}\ \emph {et~al.}(2013)\citenamefont {Serbyn},
  \citenamefont {Papi{\'{c}}},\ and\ \citenamefont {Abanin}}]{Serbyn2013}%
  \BibitemOpen
  \bibfield  {author} {\bibinfo {author} {\bibfnamefont {M.}~\bibnamefont
  {Serbyn}}, \bibinfo {author} {\bibfnamefont {Z.}~\bibnamefont {Papi{\'{c}}}},
  \ and\ \bibinfo {author} {\bibfnamefont {D.~A.}\ \bibnamefont {Abanin}},\
  }\Doi {10.1103/PhysRevLett.110.260601} {\bibfield  {journal} {\bibinfo
  {journal} {Phys. Rev. Lett.},\ }\textbf {\bibinfo {volume} {110}},\ \bibinfo
  {pages} {260601} (\bibinfo {year} {2013})},\ ISSN \bibinfo {issn}
  {0031-9007},\ \Eprint {http://arxiv.org/abs/1304.4605} {arXiv:1304.4605}
  \BibitemShut {NoStop}%
\bibitem [{\citenamefont {Kim}\ and\ \citenamefont {Huse}(2013)}]{Kim2013}%
  \BibitemOpen
  \bibfield  {author} {\bibinfo {author} {\bibfnamefont {H.}~\bibnamefont
  {Kim}}\ and\ \bibinfo {author} {\bibfnamefont {D.~A.}\ \bibnamefont {Huse}},\
  }\Doi {10.1103/PhysRevLett.111.127205} {\bibfield  {journal} {\bibinfo
  {journal} {Phys. Rev. Lett.},\ }\textbf {\bibinfo {volume} {111}},\ \bibinfo
  {pages} {127205} (\bibinfo {year} {2013})},\ ISSN \bibinfo {issn}
  {00319007},\ \Eprint {http://arxiv.org/abs/1306.4306} {arXiv:1306.4306}
  \BibitemShut {NoStop}%
\bibitem [{\citenamefont {Kim}\ \emph {et~al.}(2014)\citenamefont {Kim},
  \citenamefont {Chandran},\ and\ \citenamefont {Abanin}}]{Kim2014a}%
  \BibitemOpen
  \bibfield  {author} {\bibinfo {author} {\bibfnamefont {I.~H.}\ \bibnamefont
  {Kim}}, \bibinfo {author} {\bibfnamefont {A.}~\bibnamefont {Chandran}}, \
  and\ \bibinfo {author} {\bibfnamefont {D.~A.}\ \bibnamefont {Abanin}},\
  }\href {http://arxiv.org/abs/1412.3073} {\bibfield  {journal} {\bibinfo
  {journal} {arXiv:1412.3073}} (\bibinfo {year} {2014})},\ \Eprint
  {http://arxiv.org/abs/1412.3073} {arXiv:1412.3073} \BibitemShut {NoStop}%
\bibitem [{\citenamefont {Deng}\ \emph {et~al.}(2017)\citenamefont {Deng},
  \citenamefont {Li}, \citenamefont {Pixley}, \citenamefont {Wu},\ and\
  \citenamefont {{Das Sarma}}}]{Deng2016}%
  \BibitemOpen
  \bibfield  {author} {\bibinfo {author} {\bibfnamefont {D.-L.}\ \bibnamefont
  {Deng}}, \bibinfo {author} {\bibfnamefont {X.}~\bibnamefont {Li}}, \bibinfo
  {author} {\bibfnamefont {J.~H.}\ \bibnamefont {Pixley}}, \bibinfo {author}
  {\bibfnamefont {Y.-L.}\ \bibnamefont {Wu}}, \ and\ \bibinfo {author}
  {\bibfnamefont {S.}~\bibnamefont {{Das Sarma}}},\ }\Doi
  {10.1103/PhysRevB.95.024202} {\bibfield  {journal} {\bibinfo  {journal}
  {Phys. Rev. B},\ }\textbf {\bibinfo {volume} {95}},\ \bibinfo {pages}
  {024202} (\bibinfo {year} {2017})},\ ISSN \bibinfo {issn} {2469-9950},\
  \Eprint {http://arxiv.org/abs/1607.08611} {arXiv:1607.08611} \BibitemShut
  {NoStop}%
\bibitem [{\citenamefont {{Bar Lev}}\ \emph {et~al.}(2015)\citenamefont {{Bar
  Lev}}, \citenamefont {Cohen},\ and\ \citenamefont {Reichman}}]{BarLev2015}%
  \BibitemOpen
  \bibfield  {author} {\bibinfo {author} {\bibfnamefont {Y.}~\bibnamefont {{Bar
  Lev}}}, \bibinfo {author} {\bibfnamefont {G.}~\bibnamefont {Cohen}}, \ and\
  \bibinfo {author} {\bibfnamefont {D.~R.}\ \bibnamefont {Reichman}},\ }\Doi
  {10.1103/PhysRevLett.114.100601} {\bibfield  {journal} {\bibinfo  {journal}
  {Phys. Rev. Lett.},\ }\textbf {\bibinfo {volume} {114}},\ \bibinfo {pages}
  {100601} (\bibinfo {year} {2015})},\ ISSN \bibinfo {issn} {0031-9007},\
  \Eprint {http://arxiv.org/abs/1407.7535} {arXiv:1407.7535} \BibitemShut
  {NoStop}%
\bibitem [{\citenamefont {Agarwal}\ \emph {et~al.}(2015)\citenamefont
  {Agarwal}, \citenamefont {Gopalakrishnan}, \citenamefont {Knap},
  \citenamefont {M{\"{u}}ller},\ and\ \citenamefont {Demler}}]{Agarwal14}%
  \BibitemOpen
  \bibfield  {author} {\bibinfo {author} {\bibfnamefont {K.}~\bibnamefont
  {Agarwal}}, \bibinfo {author} {\bibfnamefont {S.}~\bibnamefont
  {Gopalakrishnan}}, \bibinfo {author} {\bibfnamefont {M.}~\bibnamefont
  {Knap}}, \bibinfo {author} {\bibfnamefont {M.}~\bibnamefont {M{\"{u}}ller}},
  \ and\ \bibinfo {author} {\bibfnamefont {E.}~\bibnamefont {Demler}},\ }\Doi
  {10.1103/PhysRevLett.114.160401} {\bibfield  {journal} {\bibinfo  {journal}
  {Phys. Rev. Lett.},\ }\textbf {\bibinfo {volume} {114}},\ \bibinfo {pages}
  {160401} (\bibinfo {year} {2015})},\ ISSN \bibinfo {issn} {10797114},\
  \Eprint {http://arxiv.org/abs/1408.3413} {arXiv:1408.3413} \BibitemShut
  {NoStop}%
\bibitem [{\citenamefont {Chandran}\ \emph {et~al.}(2016)\citenamefont
  {Chandran}, \citenamefont {Pal}, \citenamefont {Laumann},\ and\ \citenamefont
  {Scardicchio}}]{Chandran2016}%
  \BibitemOpen
  \bibfield  {author} {\bibinfo {author} {\bibfnamefont {A.}~\bibnamefont
  {Chandran}}, \bibinfo {author} {\bibfnamefont {A.}~\bibnamefont {Pal}},
  \bibinfo {author} {\bibfnamefont {C.~R.}\ \bibnamefont {Laumann}}, \ and\
  \bibinfo {author} {\bibfnamefont {A.}~\bibnamefont {Scardicchio}},\ }\Doi
  {10.1103/PhysRevB.94.144203} {\bibfield  {journal} {\bibinfo  {journal}
  {Phys. Rev. B},\ }\textbf {\bibinfo {volume} {94}},\ \bibinfo {pages}
  {144203} (\bibinfo {year} {2016})},\ ISSN \bibinfo {issn} {1550235X},\
  \Eprint {http://arxiv.org/abs/1605.00655} {arXiv:1605.00655} \BibitemShut
  {NoStop}%
\bibitem [{\citenamefont {Luitz}\ \emph {et~al.}(2016)\citenamefont {Luitz},
  \citenamefont {Laflorencie},\ and\ \citenamefont {Alet}}]{Luitz2016}%
  \BibitemOpen
  \bibfield  {author} {\bibinfo {author} {\bibfnamefont {D.~J.}\ \bibnamefont
  {Luitz}}, \bibinfo {author} {\bibfnamefont {N.}~\bibnamefont {Laflorencie}},
  \ and\ \bibinfo {author} {\bibfnamefont {F.}~\bibnamefont {Alet}},\ }\Doi
  {10.1103/PhysRevB.93.060201} {\bibfield  {journal} {\bibinfo  {journal}
  {Phys. Rev. B},\ }\textbf {\bibinfo {volume} {93}},\ \bibinfo {pages}
  {060201} (\bibinfo {year} {2016})},\ ISSN \bibinfo {issn} {2469-9950},\
  \Eprint {http://arxiv.org/abs/1511.05141} {arXiv:1511.05141} \BibitemShut
  {NoStop}%
\bibitem [{\citenamefont {Luitz}\ and\ \citenamefont {Lev}(2017)}]{Luitz2016b}%
  \BibitemOpen
  \bibfield  {author} {\bibinfo {author} {\bibfnamefont {D.~J.}\ \bibnamefont
  {Luitz}}\ and\ \bibinfo {author} {\bibfnamefont {Y.~B.}\ \bibnamefont
  {Lev}},\ }\Doi {10.1002/andp.201600350} {\bibfield  {journal} {\bibinfo
  {journal} {Ann. Phys.},\ }\textbf {\bibinfo {volume} {529}},\ \bibinfo
  {pages} {1600350} (\bibinfo {year} {2017})},\ ISSN \bibinfo {issn}
  {00033804},\ \Eprint {http://arxiv.org/abs/1610.08993} {arXiv:1610.08993}
  \BibitemShut {NoStop}%
\bibitem [{\citenamefont {Luitz}\ and\ \citenamefont {{Bar
  Lev}}(2017)}]{Luitz2017}%
  \BibitemOpen
  \bibfield  {author} {\bibinfo {author} {\bibfnamefont {D.~J.}\ \bibnamefont
  {Luitz}}\ and\ \bibinfo {author} {\bibfnamefont {Y.}~\bibnamefont {{Bar
  Lev}}},\ }\Doi {10.1103/PhysRevB.96.020406} {\bibfield  {journal} {\bibinfo
  {journal} {Phys. Rev. B},\ }\textbf {\bibinfo {volume} {96}},\ \bibinfo
  {pages} {020406} (\bibinfo {year} {2017})},\ ISSN \bibinfo {issn}
  {2469-9950},\ \Eprint {http://arxiv.org/abs/1702.03929} {arXiv:1702.03929}
  \BibitemShut {NoStop}%
\end{thebibliography}%

\end{document}